\documentclass[sigconf]{acmart} 
\renewcommand\footnotetextcopyrightpermission[1]{} % removes footnote with conference information in first column
\usepackage{balance} % for balancing columns on the final page
\usepackage{placeins}
\usepackage{float}      % provides [H] (hard placement), use only if needed
\usepackage{makecell}

\setcopyright{none}
\acmConference{}{}{}
\copyrightyear{}
\acmYear{}
\acmDOI{}
\acmPrice{}
\acmISBN{}

\acmSubmissionID{482}

\title[TriAgent: Automated Biomarker Discovery with Deep Research Grounding]{TriAgent: Automated Biomarker Discovery with Deep Research Grounding for Triage in Acute Care by LLM-Based Multi-Agent Collaboration}

\author{Kerem Delikoyun*}
\affiliation{
  \institution{Technical University of Munich}
  \city{Munich}
  \country{Germany}}
\affiliation{
  \institution{TUMCREATE}
  \city{Singapore}
  \country{Singapore}}

\author{Qianyu Chen}
\affiliation{
  \institution{Technical University of Munich}
  \city{Munich}
  \country{Germany}}
\affiliation{
  \institution{TUMCREATE}
  \city{Singapore}
  \country{Singapore}}

\author{Win Sen Kuan}
\affiliation{
  \institution{National University of Singapore}
  \city{Singapore}
  \country{Singapore}}
\affiliation{
  \institution{National University Hospital}
  \city{Singapore}
  \country{Singapore}}

\author{John Tshon Yit Soong}
\affiliation{
  \institution{National University of Singapore}
  \city{Singapore}
  \country{Singapore}}
\affiliation{
  \institution{National University Hospital}
  \city{Singapore}
  \country{Singapore}}

\author{Matthew Edward Cove}
\affiliation{
  \institution{National University Hospital}
  \city{Singapore}
  \country{Singapore}}

\author{Oliver Hayden*}
\affiliation{
  \institution{Technical University of Munich}
  \city{Munich}
  \country{Germany}}
\affiliation{
  \institution{TUMCREATE}
  \city{Singapore}
  \country{Singapore}}

\begin{abstract}
Emergency departments worldwide face rising patient volumes, workforce shortages, and variability in triage decisions that threaten the delivery of timely and accurate care. Current triage methods rely primarily on vital signs, routine laboratory values, and clinicians’ judgment, which, while effective, often miss emerging biological signals that could improve risk prediction for infection typing or antibiotic administration in acute conditions. To address this challenge, we introduce TriAgent, a large language model (LLM)-based multi-agent framework that couples automated biomarker discovery with deep research for literature-grounded validation and novelty assessment. TriAgent employs a supervisor research agent to generate research topics and delegate targeted queries to specialized sub-agents for evidence retrieval from various data sources. Findings are synthesized to classify biomarkers as either grounded in existing knowledge or flagged as novel candidates, offering transparent justification and highlighting unexplored pathways in acute care risk stratification. Unlike prior frameworks limited to existing routine clinical biomarkers, TriAgent aims to deliver an end-to-end framework from data analysis to literature grounding to improve transparency, explainability and expand the frontier of potentially actionable clinical biomarkers. Given a user’s clinical query and quantitative triage data, TriAgent achieved a topic adherence F1 score of $55.7 \pm 5.0\%$, surpassing the CoT-ReAct agent by over 10\%, and a faithfulness score of $0.42 \pm 0.39$, exceeding all baselines by more than 50\%. Across experiments, TriAgent consistently outperformed state-of-the-art LLM-based agentic frameworks in biomarker justification and literature-grounded novelty assessment. We share our repo: https://github.com/CellFace/TriAgent.
\end{abstract}


\begin{document}

\pagestyle{fancy}
\fancyhead{}

\makeatletter
% Completely disable the authors' addresses footer
\renewcommand{\@authorsaddresses}{}     % clear internal block
\makeatother

\authorsaddresses{}                     % ensure empty
\settopmatter{printacmref=false}
\renewcommand{\shortauthors}{}

\maketitle

% Manual corresponding author line under left column
\begingroup
  \renewcommand\footnoterule{}     % no horizontal rule
  \renewcommand\thefootnote{}      % no marker
  \vspace*{-1.2\baselineskip}      % adjust vertically if needed
  \noindent\parbox[t]{0.48\textwidth}{%
    \footnotesize *Corresponding authors:
    kerem.delikoyun@tum-create.edu.sg \& oliver.hayden@tum.de%
  }
\endgroup

%%%%%%%%%%%%%%%%%%%%%%%%%%%%%%%%%%%%%%%%%%%%%%%%%%%%%%%%%%%%%%%%%%%%%%%%

\section{Introduction}
Triage in acute clinical settings serves as a critical filter, precisely identifying patients who require urgent intervention and determining how resources should be allocated in overwhelmingly busy healthcare environments \cite{RN63}. Traditionally, this process hinges on well-established biomarkers: vital signs, blood tests, and structured scoring systems to define risk categories \cite{RN64}. Although these tools set a baseline for decision-making, they inherently limit clinicians to a narrow scope of diagnostic insight. Misdiagnosing infections remains a leading cause of diagnostic error in clinical practice \cite{RN82}. Individuals at risk of misdiagnosis were typically older and presented with more complex medical conditions, where heterogeneous sepsis manifestations impede prompt diagnosis \cite{RN83}. Beyond immediate stabilization, risk stratification plays a pivotal role in acute care, allowing clinicians to differentiate between patient groups with varying prognoses \cite{RN72}. For instance, rapidly distinguishing infection types directly guides therapeutic action, patients with bacterial infections should promptly receive antibiotics, while others may avoid unnecessary treatment \cite{RN73}. Similarly, robust risk prediction enables hospitals to admit only those patients at high risk of deterioration, while safely managing lower-risk individuals in outpatient settings \cite{RN74}. Such stratification not only improves patient outcomes but also alleviates the burden on already strained emergency resources. Emerging or underappreciated biomarkers may go unnoticed despite their potential to refine triage accuracy. Yet, the challenge extends beyond detection: clinicians must understand why a biomarker is pertinent, whether its relevance is backed by medical findings, and importantly, whether its novelty justifies further evaluation \cite{RN75}. A transparent, evidence-grounded rationale is essential, especially when introducing novel biomarkers into clinical workflows. Despite progress in clinical decision support, current triage systems remain constrained by limited interpretability to emerging clinical signals.
\section{Related Work}
\subsection{Biomarker Discovery with LLMs}
To overcome the limitations of biomarker discovery with traditional modeling techniques, AI has emerged as an invaluable tool \cite{RN77,RN76}. Recent works show that LLMs can actively drive biomarker discovery while grounding claims in the biomedical literature. An agentic system orchestrates modular tools (retrieval, code execution, databases) with an LLM to automate biomarker discovery and produce enrichment reports with traceable, cited evidence \cite{RN54}. In parallel, a human-augmented LLM workflow that prioritizes candidate transcriptomic biomarkers using multi-step prompts with GPT-4/Claude \cite{RN55}. Their pipeline repeatedly elevated GPX4 as a top erythroid-module candidate, illustrating how LLMs can surface plausible targets that withstand expert review and can be routed to assay development. Complementary systems for literature-based discovery (LBD) restrict LLM answers to retrieved, cited sources and integrate a user’s experimental context \cite{RN57}. 

% put this a few lines BEFORE the page where you want it to appear

\enlargethispage{\baselineskip}
\subsection{Multi-Agent Collaboration for Clinical Reasoning and Biomarker Discovery}
Beyond single-agent workflows, multi-agent systems emulate collaborative clinical reasoning, improving interpretability and grounding. Hierarchical frameworks couple role-specialized LLMs with knowledge graphs, assigning roles such as genomics, proteomics, and clinical interpretation, to dynamically contextualize putative biomarkers \cite{RN56}. General practitioner and specialist agents (genomics, proteomics, clinical interpretation) are assigned to dynamically build a medical knowledge graph to contextualize and cite putative biomarkers. Multi-Agent Conversation (MAC) framework, where three doctor agents and a supervisory agent, all powered by GPT-4, engaged in interactive diagnosis of rare diseases, outperforming single-model and chain-of-thought \cite{RN69} prompting approaches \cite{RN60}. Moreover, MedAgents framework  demonstrates how role-playing agents engage in expert gathering, individual analyses, report summarization, collaborative consultation, and decision-making, simulating multidisciplinary rounds in clinical practice \cite{RN65}. This training-free setup enhances reasoning and interpretability without requiring domain-specific fine-tuning, achieving state-of-the-art performance across various medical benchmarks. Such findings illustrate how structured multi-agent collaboration can surface domain expertise otherwise inaccessible through single agent architectures. Other adaptive agent systems tailor complexity dynamically: one framework assembles specialist or generalist agents depending on case difficulty \cite{RN61}. Similarly, role-specific LLM agents simulate emergency department staff, using self-confidence metrics and retrieval-augmented generation based on the ESI handbook \cite{RN53}. This system significantly improved triage classification and closed the gap to human-level accuracy in real-world clinical datasets. Beyond triage, modular multi agent framework for ICU decision support deploys specialized agents for lab results, vital interpretation, and contextual reasoning, with a validation agent ensuring ethical oversight and transparency \cite{RN62}.
\enlargethispage{\baselineskip}
\subsection{Problem Statement and Our Contribution}
Despite these advances, no prior work couples automated biomarker discovery with literature-grounded deep research for novelty assessment in acute care triage. Existing multi-agent triage frameworks largely focus on following established guidelines with clinically known biomarkers for guideline-aware decisions with retrieval-augmented evidence in emergency settings by leveraging LLM-based single or multi-agent collaborations. Therefore, these systems currently aim to operate and orchestrate the existing clinical protocols rather than discovering novel biomarkers which could potentially provide faster response time and lower cost with higher specificity for acute-care triage. These constraints point to four unmet challenges: i) traditional frameworks do not surface novel biomarkers dynamically, ii) literature grounding is often absent, iii) practical clinical translation is hindered by analytical and workflow variability, and iv) clinical utility remains uncertain without context-specific validation. 

Addressing these gaps requires a unified, explainable system that not only discovers novel biomarkers from acute care clinical patient data but also rigorously validates them against existing biomedical knowledge and shown to be effective on real-world clinical scenarios for triage decisions. We introduce TriAgent, an LLM-based multi-agentic framework built to discover, ground, and explain novel biomarkers poised to improve acute care triage. At its core, TriAgent consists of two complementary systems: i) a data analysis agent that autonomously mines patient data readily available from routine tests to identify candidate biomarkers not traditionally used in triage and ii) a deep research agent for literature-grounding that retrieves and consolidates supporting medical evidence. If the biomarker appears in literature, it is marked as grounded. If not, TriAgent flags it as novel for validation, and generates a structured justification on biological patterns, and reasoning to explain clinical significance.

\begin{figure*}[!t]
  \centering
  \includegraphics[width=\textwidth]{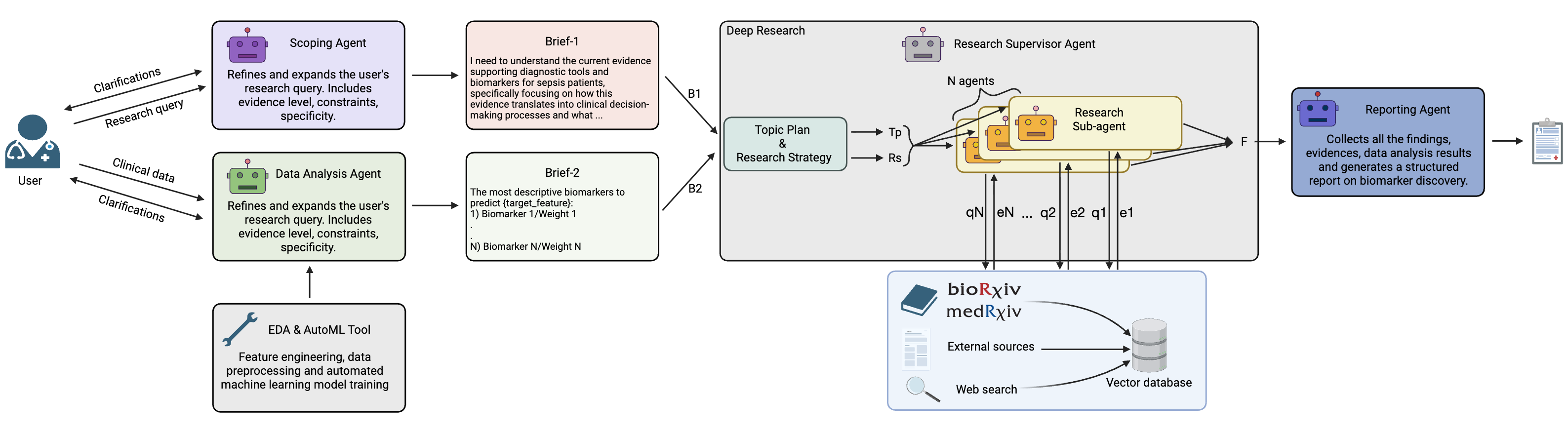}
  \caption{Architecture of TriAgent. TriAgent is a graph-based multi-agent framework for automated biomarker discovery and literature-grounded validation. The Scoping and Data Analysis Agents produce structured briefs (Brief-1, Brief-2); the Research Supervisor Agent coordinates sub-agents for RAG-based searches; the Reporting Agent integrates findings into an auditable report.}
  \label{fig:triagent-arch}
  \Description{Two-column architecture diagram as described in the caption.}
\end{figure*}

\enlargethispage{\baselineskip}
\section{Methodology}
TriAgent is a graph-based multi-agent system designed for automated biomarker discovery and literature-based validation and justification. The graph flow coordinates specialized agents through a structured workflow to transform raw data and research queries into auditable findings that distinguish grounded biomarkers from novel candidates. TriAgent follows a multi-stage pipeline for scoping, data analysis, research supervision, topic execution, and reporting, each stage handled by dedicated agents and tools.

The Scoping Agent is responsible for refining the user’s initial research query. It engages in clarification dialogues to establish the problem statement, success criteria, and specifics. The outcome of this stage is called Brief-1. Formally, given a query, the scoping agent generates a structured research brief, $B_1 = f_s(Q, C)$, where $C$ represents user-defined constraints (e.g., evidence level, patient demographics, medical conditions/comorbidity, etc.). This step ensures that the downstream analysis is well aligned with the clinical or scientific objectives.

The Data Analysis Agent integrates exploratory and predictive modeling to identify potential biomarkers. It orchestrates the exploratory data analysis (EDA) tool to summarize data distributions and generate descriptive statistics, S. It then calls the AutoML pipeline to perform preprocessing (e.g., scaling, normalization, autotask detection, random train/valid/test set splitting), hyperparameter tuning, model training and selection, and explainability analysis (e.g., feature importance, SHAP values). Given dataset, $D$, target, $y$, and configuration, $\theta$, the AutoML pipeline outputs the best model, $M^{*}$, along with the importance of the features, $\phi_i$, for each candidate biomarker, $x_i^{*}$:
\begin{equation}
f_{\text{automl}}(D, y, \theta) = M^{*}, \{\phi_i, x_i^{*}\}
\end{equation}
Feature importance scores, $\phi_i$, are normalized such that:
\begin{equation}
\sum_{i=1}^{n} \phi_i = 1
\end{equation}
The synthesized output, Brief-2, $B_2$, is structured as follows:
\begin{equation}
B_2 = \{(x_i^{*}, \phi_i) \mid i = 1, \ldots, n\}
\end{equation}
Following automated data analysis, the synthesized output, Brief-2, includes most descriptive biomarkers (n = 10) with importance weights. Importantly, each biomarker is linked to an evidence hook that seeds subsequent literature probing. 

The Research Supervisor Agent coordinates the deep research by transitioning insights from refined user query and data analysis to literature research. Consuming Brief-1 and Brief-2, it formulates a topic plan, $T_p$, a set of topic-specific queries, $q_i$, and strategies to investigate each candidate biomarker, $x_i{*}$, and a research strategy, $R_s$, specifying thresholds for grounding, $\tau_g$, and novelty, $\nu$. The research supervisor agent then uses thinking tool to decide if multiple agents should be deployed and if so, it uses conduct research tool to spawn multiple research sub-agents to conduct topic-specific literature research.
\begin{equation}
T_p = \mathrm{LLM}(B_1, B_2) \rightarrow q_1, q_2, q_3, \ldots, q_n\ \&\ R_s = (\tau_g, \nu)
\end{equation}
Each research sub-agent executes targeted literature retrieval for a research topic, $q_i$, associated with a candidate biomarker, $x_i{*}$. Using the retrieval augmented generation (RAG) tool, the agents perform multi-source searches across pre-indexed biomedical corpora (e.g., Biorxiv, Medrxiv), web sources and optional external data sources provided by user. Given research topic, $q_i$, each retrieved evidence fragment, $e_{ij}$, is assigned a similarity score:
\begin{equation}
s_{ij} = \cos(\vec{q}_i, \vec{e}_{ij})
\end{equation}
All the retrieved evidence fragments are re-ranked before returned to Supervisor Research Agent which decides whether to conduct further research or to terminate the deep research for particular research topic. Upon completion of topic research, retrieved findings, $F = \{ f_i \mid i = 1, \ldots, n \},$ are assigned evidence IDs, $E = \{ e_i \mid i = 1, \ldots, n \},$ for traceability. The Research Supervisor Agent aggregates findings, {F}, from all research sub-agents, deduplicates claims, and merges evidence trails to decide biomarkers can be classified as grounded given research strategy if sufficient and diverse literature evidence are found, or novel if retrieval consistently fails to establish support. Novelty confidence is strengthened by systematic negative searches, query refinement, and coverage of multiple databases. This stage also produces gap analyses to highlight areas where evidence remains inconclusive.

The Reporting Agent synthesizes the full pipeline outputs into a structured and detailed report, {R}. The report includes classification of biomarkers as grounded or novel. Evidence trails with citations (e.g., DOI or URLs), justifications and novelty confidence scores, analysis results and plots of EDA/AutoML tools for explainability and finally, limitations and gaps-in-evidence annotations. This ensures clinicians and researchers receive not only candidate biomarkers but also an auditable rationale for their inclusion or novelty claims.
\begin{equation}
R = \mathrm{LLM}(B_1, B_2, F, E)
\end{equation}
\section{Experimental Setup}
\subsection{Dataset}
The dataset used in this study was curated from two well-defined patient cohorts: a control group of day-surgery patients (n = 20) serving as healthy reference samples and a fever cohort of emergency department (ED) patients (n = 94) presenting with acute infectious conditions. Inclusion criteria were defined across cohorts: for the fever group, only adults over 21 years of age with fever of body temperature exceeding 38.3°C within 24 hours of ED arrival were included; and for the control group, only adults over 21 years undergoing elective day-case surgery without any symptoms or signs of infection and with a body temperature below 37.5°C were included. For all cohorts, the existence and type of infection were confirmed through the hospital’s microbiology department records. For each encounter, demographic information such as age and sex, physiological parameters such as heart rate, respiratory rate routinely collected at or before triage, and complete blood count (CBC) panels with differential (neutrophils, lymphocytes, monocytes, eosinophils, basophils) were included.

From these raw measurement values, engineered features capturing immune and inflammatory responses were derived using the exploratory data analysis (EDA) tool. The primary targets for supervised learning were infection type (bacterial, viral, or other). All data were automatically checked for missingness, normalized, and partitioned into training, validation, and test sets using the AutoML tool. All blood samples and clinical records were fully de-identified, with study conduct approved under Domain Specific Institutional Review Board protocols from the National University Hospital, Singapore, in accordance with the Domestic Study Review Board and the Declaration of Helsinki. The details and composition of clinical dataset is provided in Appendix A.

\subsection{Implementation}
The TriAgent framework is designed as a graph of agents and tools implemented in LangGraph and LangChain libraries, where each node corresponds to a functional component and edges define the execution order. Each edge can be direct, enforcing sequential execution, or conditional, allowing transitions only when specified thresholds are met. The temperature of different agents was set for more deterministic grounding/justification (T=0) with the cache seed of 42 for reproducibility. The maximum number of iterations were set as 10 and number of research sub-agents as 5. We deployed and compared the following LLM models using either direct API calls from the LLM provider: OpenAI GPT-4o or through Amazon Web Services (AWS) Bedrock: Anthropic Sonnet 4, OpenAI GPT OSS 20B.

\subsection{Baselines}
\begin{itemize}
    \item Reason and Act (ReAct) agent \cite{RN68} combines verbal reasoning traces with actions (e.g. search, tool invocation), interleaving reasoning and external tool use. As a baseline, a single agent relying on ReAct is used to execute both the data analysis and literature search and grounding/justification steps. ReAct agents are prompted with vanilla, Chain-of-thought and Self consistency with tool use capability for fair comparison against TriAgent.
    \item Chain-of-thought (CoT) reasoning \cite{RN69} was used as a baseline by prompting LLMs with explicit step-by-step instructions to simulate structured reasoning when applied to deep research. CoT was implemented by directly appending reasoning directives to the prompt and measuring whether structured reasoning improved consistency in biomarker grounding/justification.
    \item Self-consistency (SC) \cite{RN70} was included to generate multiple reasoning chains for the same query and aggregates the responses by majority voting, thereby reducing variance in LLM outputs. We applied self-consistency to research queries, evaluating whether aggregating reasoning paths led to more reliable grounding/justification through deep research.
\end{itemize}

\subsection{Evaluation Protocol}
TriAgent is evaluated along multiple complementary dimensions to understand not only how accurately it grounds and justifies biomarkers, but also how well it adheres to domain constraints and factuality. Our first evaluation metric is topic adherence which is assessed by measuring whether the outputs of the deep research and retrieval stages remain within the predefined scope established by the user’s research query/goal. Second, we evaluate the agent’s consistency and reliability in factuality by measuring faithfulness, indicating that given the retrieved context (e.g., literature search) how factually consistent the response is. And finally, we compare TriAgent’s full pipeline containing Research Supervisor Agent coordinating multiple research sub-agents against a baseline in which a ReAct agent is prompted through vanilla, CoT and SC. We examine differences in grounding and novelty judgement and coverage of supporting literature.
\subsection{Metrics}
We measure whether TriAgent stays within the domains fixed by user’s research query/goal by measuring topic adherence score. The metric takes a transcript including the queries and topics that define the allowed scope to determine whether the agent’s answer adheres to any of the specified topics. This directly operationalizes “on-topic” behavior during retrieval and justification. To quantify topic adherence following metrics were calculated: Precision measures, among all answered queries, the fraction whose answers adhere to at least one of the reference topics (penalizing off-topic answers that the agent nonetheless attempted). Recall measures coverage of on-topic questions by counting how often the agent answered topically when it should have answered, penalizing inappropriate refusals on in-scope queries. F1 score is the harmonic mean of precision and recall and summarizes adherence as a single score and thus, F1 score is reported in this work for topic adherence.

The second metric we used is faithfulness which measures how factually consistent a response is with the retrieved context. Faithfulness is calculated based on the whether the claims in the response supported by the retrieved context compared to total claims in the response. 
\begin{equation}
\textit{Precision} = \frac{N_{TA}}{N_{TA} + N_{\overline{TA}}}
\end{equation}

\begin{equation}
\textit{Recall} = \frac{N_{TA}}{N_{TA} + N_{\text{should}}}
\end{equation}

\begin{equation}
F1 = \frac{2 \times \textit{Precision} \times \textit{Recall}}{\textit{Precision} + \textit{Recall}}
\end{equation}

\begin{equation}
\textit{Faithfulness} = \frac{C_S}{C_T}
\end{equation}
where ${N_{TA}}$ is number of queries (or retrieval/generation instances) that are answered and adhere to at least one reference topic. $N_{T\overline{A}}$ is number of queries that are answered but do not adhere to any reference topic. $N_{should}$ is number of queries that should have been answered (they are in-scope given reference topics) but were refused or answered off-topic. On the other hand, $C_S$ is the number of claims in the response supported by the retrieved context and $C_T$ is the total number of claims in the response.
\section{Results}
The performance of TriAgent is evaluated using multiple foundation models and against baseline methods. The results reported in this section are averaged over three repeated runs. Table 1 represents the overall results indicating the comparison of TriAgent against baseline methods for different tasks (i.e., ReAct prompted by vanilla LLM, CoT, SC). In this section, we present an ablation study examining optimal sub-agent configuration and the influence of foundation models. Following the evaluation of TriAgent with time and cost analyses.

\begin{table}[t]
\centering
\caption{Performance comparison of \textbf{TriAgent} against various baseline methods on the grounding of biomarker discovery on topic adherence and faithfulness given Brief-1 and Brief-2. Higher topic adherence (F1-score) and faithfulness values indicate stronger performance. Values are given as three repeated average~$\pm$~standard deviation.}
\label{tab:performance}
\renewcommand{\arraystretch}{1.2}
\small
\begin{tabular}{lcc}
\toprule
& \multicolumn{2}{c}{\textbf{Topic Adherence (F1) / Faithfulness}} \\ % <-- Centered header
\cmidrule(lr){2-3}
\textbf{Method} & \textbf{Sonnet 4} & \textbf{GPT-4o} \\
\midrule
\multicolumn{3}{l}{\textbf{ReAct Agents}} \\
Vanilla & $48.1 \pm 10.5$\% / $0.03 \pm 0.06$ & $35.0 \pm 5.8$\% / $0.16 \pm 0.17$ \\
w/CoT   & $50.0 \pm 10.0$\% / $0.22 \pm 0.20$ & $\boldsymbol{\mathit{50.4 \pm 14.3\%}} / 0.21 \pm 0.10$ \\
w/SC    & $42.9 \pm 2.5$\% / $0.24 \pm 0.25$  & $47.8 \pm 5.8$\% / $0.56 \pm 0.09$ \\
\midrule
\textbf{TriAgent} & $\boldsymbol{\mathit{55.7 \pm 5.0\%} / \mathit{0.42 \pm 0.39}}$ & $36.5 \pm 7.9\% /  \boldsymbol{\mathit{0.68 \pm 0.13}}$ \\
\bottomrule
\end{tabular}
\end{table}

\subsection{Performance Comparison with Baseline Methods}
As a reference point, we first evaluated the performance of a vanilla ReAct agent without any auxiliary reasoning strategies or multi-agent orchestration. In this setting, the LLM was prompted directly with the user’s research query with identified biomarker candidates. Results show that a standalone LLM achieved fairly high topic adherence score of $48.1 \pm 10.5\%$ while the lowest faithfulness score among all agents ($0.03 \pm 0.06\%$), indicating that it managed to remain within the predefined domain, however, often produced incomplete evidence trails, and lacks systematic novelty assessment.

While CoT encourages stepwise reasoning, it often produces verbose chains prone to drift outside the target domain and introduces hallucinations when handling complex terminologies and topics as faithfulness score indicates. As topic plan generated by research supervisor agent includes a research strategy for sub-agent to perform the deep research, TriAgent mitigates the risk of overloading one LLM with too intense and overlapping concepts to pursue. Therefore, TriAgent’s multi-agent structure constrains reasoning within scoped briefs and supervised topics, resulting in more stable outputs and higher adherence to clinical domains. On average, CoT achieved $50 \pm 10\%$ in topic adherence and $0.22 \pm 0.20$ in faithfulness, approximately $10\%$ and $50\%$ lower than TriAgent, respectively.

SC could only improve the faithfulness compared to baseline performance while achieved lower in topic adherence with remarkably longer and costly reasoning path. However, it remains limited by its reliance on majority voting without structured supervision or evidence traceability. Our results show that SC achieved $42.9 \pm 2.5\%$ in topic adherence and $0.24 \pm 0.25$ in faithfulness.

The above baseline shows the performance focusing on the capabilities on literature search and grounding while adhering to the purpose and constraints of the research query. ReAct agents perform reasonably well in tool usage, retrieval and justification but lack the explicit division of roles that enables TriAgent to manage complex literature search for grounding through deep research. Without the orchestration of Research Supervisor Agent for deep research, ReAct agent tends to overlap queries, miss relevant evidence, and provide weaker novelty assessments. TriAgent’s Supervisor Research Agent/sub-agent interaction achieved broader literature retrieval and stronger evidence trail traceability. When Sonnet 4 used, TriAgent achieved $55.7 \pm 5.0\%$ in topic adherence, it also outperformed in faithfulness ($0.42 \pm 0.39$) which measures how factually consistent research supervisor and sub-agents’ responses are with the retrieved context, indicating higher consistency in literature research and grounding with novelty assessment. For GPT-4o, ReAct agent with CoT prompting achieved the highest topic adherence of $50.4 \pm 14.3\%$ however, TriAgent still achieved significantly higher faithfulness score of $0.68 \pm 0.13$.

\begin{figure}[h]
  \centering
  \includegraphics[width=0.75\linewidth]{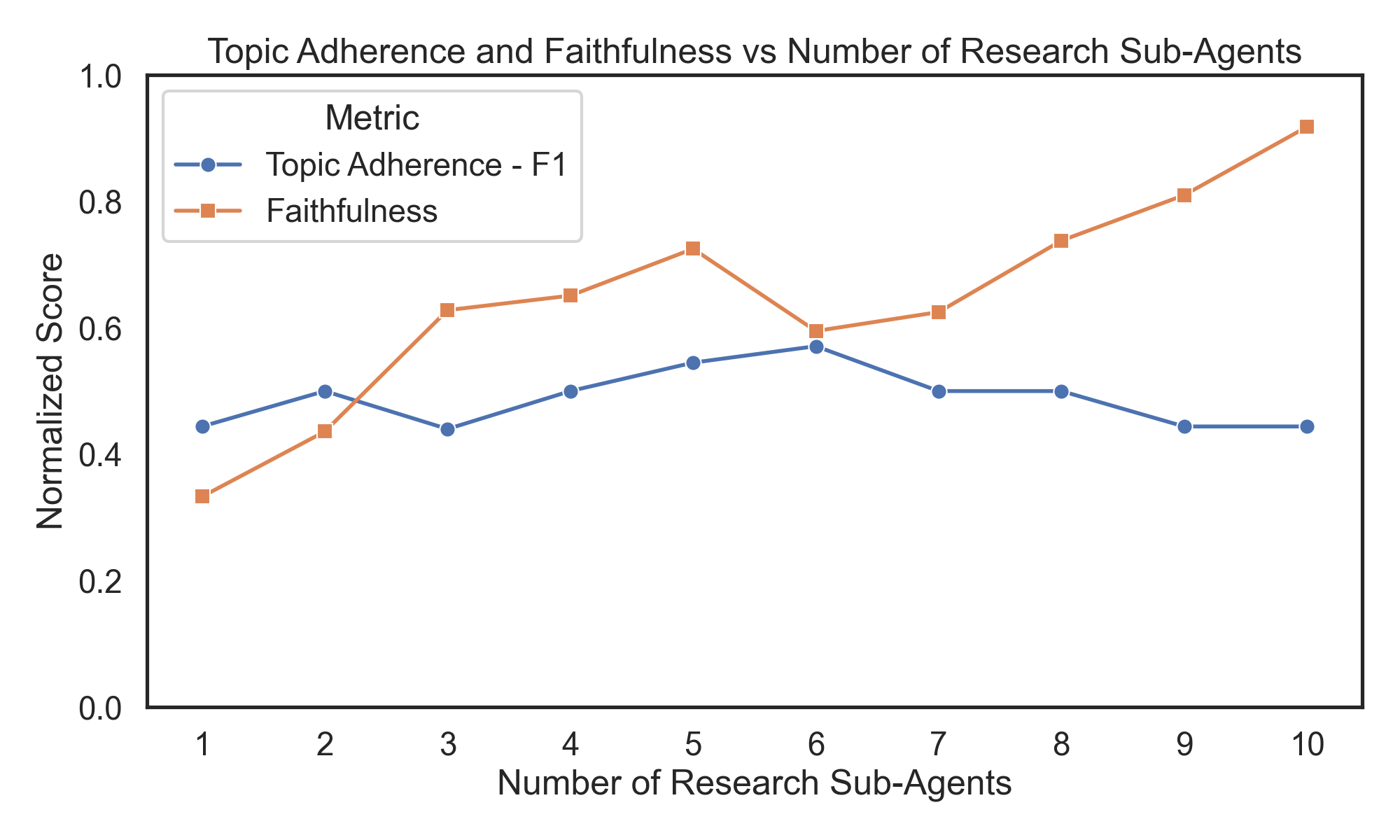}
  \caption{Evaluation of the effect of varying the number of research sub-agents on grounding quality and novelty assessment with the performance.}
  \label{fig:logo}
  \Description{Figure 2: Evaluation of the effect of varying the number of research sub-agents on grounding quality and novelty assessment with the performance.}
\end{figure}

\subsection{Ablation Study}
\textbf{Optimal Sub-agent Configuration for Deep Research}\newline
To investigate how the number of sub-agents influences performance of grounding and novelty assessment, measured in topic adherence and faithfulness metrics. We varied the configuration of the Research Supervisor Agent through its prompt, which generates topic plans and assigns subtopics to each sub-agent with research strategy. When too many sub-agents are spawned ($> 5$), topic overlap increases, leading to redundant retrievals and decreased efficiency without significant gains that leads drifting from user’s research query measured by topic adherence while faithfulness which measures factual consistency increases. Conversely, when too few sub-agents ($\leq$ 3) are deployed, individual agents must address broad or mixed subtopics, often diluting retrieval focus and reducing the precision of evidence aggregation. This can be seen in decreasing topic adherence and faithfulness. Our results indicate that there exists an optimal range of sub-agents ($\sim$4-6) where redundancy is minimized and topic specialization is preserved. However, given the execution time and cost in API call/token usage, the optimization of the resources should be also considered. Therefore, we chose 5 sub-agents as the optimal configuration, considering the trade-off between performance and cost. Within this configuration, TriAgent achieves consistently higher topic adherence and faithfulness, enabling reliable novelty assessment while maintaining a reasonable computational cost (Figure 2).\newline

\enlargethispage{\baselineskip}
\noindent\textbf{Foundation Model Comparison.}
To assess the influence of backbone LLMs, we ran TriAgent with Anthropic Sonnet 4 and OpenAI GPT-4o, integrating each into the same graph flow under identical experimental conditions. We evaluated single model runs, where the same model powered all agent roles such as scoping, research supervisor, or sub-agents. Overall, results indicate that TriAgent maintains consistent performance with single model runs providing stable outputs. However, Sonnet 4 showed higher topic adherence while lower factual consistency compared to GPT-4o on deep research (Table 2). For completeness, we also evaluated TriAgent with an open-source model GPT-OSS-20B that the comparative results are provided in Table 5 of the Appendix B.\newline

\begin{table}[t]
\centering
\caption{Evaluation of \textbf{TriAgent}'s performance of different stages in topic adherence, runtime, and cost across two foundation models. Values are given as three repeated average~$\pm$~standard deviation.}
\label{tab:evaluation}
\renewcommand{\arraystretch}{1.25}
\small
\begin{tabular}{lcc}
\toprule
\textbf{Method} & \textbf{Sonnet 4} & \textbf{GPT-4o} \\
\midrule
Topic Adherence & $55.7 \pm 5.0$\% & $36.5 \pm 7.9$\% \\
Faithfulness    & $0.42 \pm 0.39$ & $0.68 \pm 0.13$ \\
Time (s)        & $500 \pm 171$ & $508 \pm 203$ \\
Cost (Total Token) & $60,512 \pm 5,714$ & $80,045 \pm 8,050$ \\
\bottomrule
\end{tabular}
\end{table}

\noindent\textbf{Time Analysis.}
We measured the runtime efficiency of TriAgent across different models and configurations. Runtime was recorded as the total time required to complete the entire task. Results show that both tested models completed the entire task (i.e., data analysis with deep research) around 500 s. The majority of the execution time was allocated to deep research stage, where the number of sub-agents and breadth of retrieval directly influenced runtime. Table 2 presents the average run times across two foundation model.\newline

\noindent\textbf{Cost Analysis.}
We also evaluated the cost of TriAgent in terms of token usage across different foundation models. The most significant cost drivers were the deep research stage, particularly when multiple sub-agents were spawned. As presented before, Sonnet 4 reached higher topic adherence while GPT-4o showed significant gain in faithfulness. When it comes to average total token, GPT-4o showed roughly $30\%$ more consumption. Table 2 summarizes the token usage across different foundation models.\newline

\noindent\textbf{Case Study.}
A detailed case study demonstrating the end-to-end execution of TriAgent is provided in the Appendix C. It illustrates how a user query is refined by the scoping agent, how candidate biomarkers predictive of infection subtype are extracted through data analysis, and how the deep research module coordinates sub-agents to perform literature retrieval, grounding, and novelty assessment, followed by the detailed report generated by the reporting agent.

\section{Conclusions}
In this work, we introduced TriAgent, a multi-agent framework for automated biomarker discovery and literature-grounded validation through deep research. By decomposing the pipeline into dedicated agents for scoping, data analysis, deep research supervision, and reporting, TriAgent enables transparent and rigorous identification of grounded and novel biomarker candidates. Experimental evaluation shows that the architecture improves evidence coverage, reduces drift in domain adherence, and yields more stable biomarker grounding and justification compared to conventional methods such as ReAct agents with CoT and SC prompting. Future work will focus on further integration of quantitative biomarker validation for clinical use, enhancing semantic understanding to reduce evidence mis-retrieval, and balancing retrieval breadth with specificity. In summary, automated biomarker discovery through analytical AI augmented by agentic AI frameworks hold the promise of opening pathways for improved acute care triage.

\section{Limitations}
While TriAgent demonstrates promising capabilities in automating biomarker discovery and grounding novelty claims through deep research, there are several limitations to note.
\begin{itemize}
    \item Establishing evidence levels for identification of biomarkers: This work does not aim to perform clinical validation of biomarkers at the level required for regulatory or clinical use. According to biomarker qualification frameworks \cite{RN80,RN67}, reliable biomarkers must undergo analytical validation to confirm reproducibility, accuracy, sensitivity, and specificity; clinical validation to demonstrate consistent association with outcomes in adequately designed cohorts and qualification to define the context of use and decision-making utility \cite{RN81}. TriAgent can automate data analysis and assess literature support and novelty but does not yet conduct prospective medical studies or formal biomarker test verification.
    \item Data source and validation from multi centers: The data used for training and discovery in this work comes from a single healthcare facility. Patient populations and profiles with data collection practices can vary substantially across institutions. This can lead to local biases, and reduced generalizability for biomarker discovery through data analysis \cite{RN79, RN78}. To obtain broader and more reliable insights, future work should aggregate datasets across multiple institutions.
    \item Data access and availability: Even though our architecture is modular and reproducible, the performance of the deep research pipeline depends on available medical corpora, the quality of retrieved data sources and the coverage of literature limited by private data sources or access restriction by paywalls. In medical domains/conditions with sparse published evidence, the system may underperform in grounding or novelty detection through deep research due to limited access of data.
\end{itemize}

\section*{Acknowledgements}
This research is supported by the National Research Foundation, Prime Minister’s Office, Singapore, under Intra-CREATE Thematic Grant (NRF2019-THE002-0008). The figures were partly generated using BioRender (\url{https://www.biorender.com}).

\bibliographystyle{ACM-Reference-Format}
%\bibliography{references}

\clearpage         % <-- start a new page (two-column) at the left-top column
\appendix
\section{Dataset Composition}
\FloatBarrier

Tables 3 and 4 summarize the clinical data used in this work. Table 3 provides descriptive statistics of patient demographics, infection characteristics, antibiotic administration, and clinical outcomes across the control and fever cohorts, including measures such as age, sex distribution, infection type, and length of stay. Table 4 lists the numerical and categorical features extracted from the same datasets, covering vital signs, complete blood count parameters, and infection-related indicators, all of which were processed by the exploratory data analysis (EDA) and AutoML tools within the TriAgent framework to identify predictive biomarkers for infection subtyping.

%\columnbreak

\begin{table}[H]
\centering
\caption{The descriptive statistics of clinical patient data between control and fever cohorts.}
\label{tab:clinicalstats}
\renewcommand{\arraystretch}{1.2}
\small
\setlength{\tabcolsep}{3pt} % optional: slightly narrower columns
\begin{tabular}{llcc}
\toprule
\textbf{Characteristics} & & \textbf{Mean} & \textbf{Population (\%)} \\
\midrule
\multicolumn{4}{c}{\textbf{Control / Fever}} \\[0.25em] % <-- centered heading
\textbf{Age (years)} & & 57.3 / 54.0 &  \\[0.25em]
\textbf{Sex} & Female &  & 62.5 / 36.5 \\
             & Male   &  & 37.5 / 63.5 \\[0.25em]
\textbf{Infection} & Not present &  & 100 / 12 \\
                   & Bacterial    &  & 0 / 27.4 \\
                   & Viral        &  & 0 / 21.9 \\
                   & Others       &  & 0 / 38.4 \\[0.25em]
\textbf{Antibiotics} & &  & 0 / 49.5 \\
\midrule
\multicolumn{4}{l}{\textbf{Outcome}} \\[0.25em]
\textbf{Disposition} & Home         &  & 100 / 26 \\
                     & General Ward &  & 0 / 65.1 \\[0.25em]
\makecell[l]{\textbf{Hospital Length of Stay}\\\textbf{(min)}} & & 0 / 70 &  \\[0.25em]
\makecell[l]{\textbf{Intubation Duration}\\\textbf{(min)}} & & 0 / 0 &  \\
\bottomrule
\end{tabular}
\end{table}

\begin{table}[t]
\centering
\caption{The numerical and categorical features extracted from clinical patient data used by EDA and AutoML tools orchestrated by the data analysis agent in the TriAgent framework.}
\label{tab:features}
\small
\renewcommand{\arraystretch}{1.05}
\setlength{\tabcolsep}{4pt}

% ---- Numerical Features ----
\begin{tabular}{p{0.45\linewidth} p{0.45\linewidth}}
\toprule
\multicolumn{2}{c}{\textbf{Numerical Features}} \\
\midrule
1) height                     & 20) lymph($10^{3}$/uL) \\
2) weight                     & 21) mono($10^{3}$/uL) \\
3) t0\_temp\_avg              & 22) eo($10^{3}$/uL) \\
4) t0\_hr\_avg                & 23) baso($10^{3}$/uL) \\
5) t0\_rr\_avg                & 24) neut(\%) \\
6) t0\_sbp\_avg               & 25) lymph(\%) \\
7) t0\_map\_avg               & 26) mono(\%) \\
8) t0\_spo2\_avg              & 27) eo(\%) \\
9) t0\_fio2\_avg              & 28) baso(\%) \\
10) wbc($10^{3}$/uL)          & 29) ig($10^{3}$/uL) \\
11) rbc($10^{6}$/uL)          & 30) ig(\%) \\
12) hgb(g/dL)                 & 31) ret(\%) \\
13) hct(\%)                   & 32) ret($10^{6}$/uL) \\
14) mcv(fL)                   & 33) irf(\%) \\
15) mch(pg)                   & 34) ipf(\%) \\
16) mchc(g/dL)                & 35) ig\_total\_neu\_ratio(\%) \\
17) plt($10^{3}$/uL)          & 36) left\_shift\_flag \\
18) pct(\%)                   & 37) intubation\_duration \\
19) neut($10^{3}$/uL)         & 38) disposal \\
\bottomrule
\end{tabular}

\vspace{0.5em}

% ---- Categorical Features ----
\begin{tabular}{p{0.45\linewidth} p{0.45\linewidth}}
\toprule
\multicolumn{2}{c}{\textbf{Categorical Features}} \\
\midrule
39) age       & 44) infection\_type \\
40) gender    & 45) abx\_appropriate \\
41) ethnicity & 46) infect\_source \\
42) t0\_abx   & 47) hosp\_los \\
43) infection &  \\
\bottomrule
\end{tabular}
\end{table}

\FloatBarrier

\section{Results of an Open-Source Model}
In addition to the results obtained with Sonnet 4 and GPT-4o, we evaluated TriAgent using an open-source model, GPT-OSS-20B (Table 5). Compared to Sonnet 4 and GPT-4o, topic adherence showed a lower performance as $33.1 \pm 0.1\%$ while it achieved slightly higher score for faithfulness against Sonnet 4 ($0.42 \pm 0.39\%$) even though GPT-4o achieved more than $40\%$ higher faithfulness score of $0.68 \pm 0.13\%$. This may imply that even though GPT-OSS-20B is significantly smaller, it can provide factually consistent responses while it suffers from some drift against user’s original research query.

\begin{table}[H]
  \centering
  \caption{Performance of GPT-OSS-20B for the same data analysis and deep research task.}
  \label{tab:gptoss}
  \renewcommand{\arraystretch}{1.2}
  \small
  \begin{tabular}{lcc}
  \toprule
  \textbf{Method} & \textbf{Topic Adherence (F1)} & \textbf{Faithfulness} \\
  \midrule
  \textbf{TriAgent} & $33.1 \pm 0.1$\% & $0.48 \pm 0.29$ \\
  \bottomrule
  \end{tabular}
\end{table}

\FloatBarrier  

\section{Case Study}
As a case study, we illustrate the end-to-end operation of TriAgent on a clinical research query. The user begins by submitting a query, which is refined through clarification steps by the Scoping Agent into a structured research brief, Brief-1. Clinical data is then provided, and the Data Analysis Agent clarifies the training and target feature as infection sub-type to conduct exploratory data analysis and applies AutoML pipeline to extract features predictive of infection sub-type. These results are synthesized into Brief-2, which highlights candidate biomarkers with predictive power for distinguishing bacterial infections where antibiotic treatment would be appropriate. Brief-1 and Brief-2 are passed to the deep research module, where the Research Supervisor Agent generates subtopics and assigns them to sub-agents for literature search, grounding and novelty assessment. This process produces evidence trails that ground the predictive features or identify them as novel. Finally, Reporting Agent collects all the input from user, brief statements, findings and evidence to generate a report for biomarker justification with explainability.\newline

\noindent\textbf{User’s Research Query:} 
I want to investigate whether the sequential measurements of CBC-based features (neutrophil \%, eosinophil \%, and immature granulocyte \%) together with vital-sign trends serve as real-time indicators of antibiotic treatment efficacy and clinical improvement in bacterial infections.\newline

\noindent\textbf{Clarification Questions by Scoping Agent:}
To provide you with the most relevant research, I need to clarify a few key aspects of your investigation:\newline
1) Patient population: Are you focusing on a specific patient group (e.g., hospitalized patients, ICU patients, specific age groups, immunocompromised patients)?\newline
2) Bacterial infection types: Are you interested in specific types of bacterial infections (e.g., sepsis, pneumonia, UTIs, skin/soft tissue infections) or all bacterial infections generally?\newline
3) Vital signs: Which specific vital signs are you most interested in tracking alongside the CBC parameters (e.g., temperature, heart rate, blood pressure, respiratory rate, oxygen saturation)?\newline
4) Timeline: What timeframe for sequential measurements are you considering (e.g., daily measurements over 3-7 days, twice-daily measurements, specific intervals)?\newline
5) Clinical outcomes: What specific measures of "clinical improvement" are you most interested in (e.g., length of stay, mortality, symptom resolution, biomarker normalization)?\newline

\noindent\textbf{Clarification Answers by User:}\newline
1. Patient population: Hospitalized adult patients in general wards or ICUs receiving systemic antibiotics for confirmed or suspected bacterial infections, including immunocompromised cases analyzed separately for sensitivity.\newline
2. Bacterial infection types: Focus on common infections such as sepsis, bacteremia, community-acquired pneumonia, UTIs, and intra-abdominal infections, with other localized infections included if sequential data exist.\newline
3. Vital signs: Track temperature, heart rate, respiratory rate, mean arterial pressure, and oxygen saturation as indicators of inflammatory and hemodynamic response to antibiotics.\newline
4. Timeline for sequential measurements: Analyze daily CBC data and 6-12-hour vital recordings over the first 3-7 days of antibiotic therapy to capture early treatment response. 5. Clinical outcomes: Define improvement by normalization of vitals and CBC indices, shorter hospital stay, reduced ICU transfers, clinical resolution, and avoidance of antibiotic escalation.\newline

\noindent\textbf{Brief-1 by Scoping Agent:}
I want to investigate whether sequential measurements of CBC-based features (specifically neutrophil percentage, eosinophil percentage, and immature granulocyte percentage) combined with vital sign trends (temperature, heart rate, respiratory rate, mean arterial pressure, and oxygen saturation) can serve as real-time indicators of antibiotic treatment efficacy and clinical improvement in hospitalized adult patients with bacterial infections. The study should focus on patients in general wards or ICUs receiving systemic antibiotics for confirmed or suspected bacterial infections, including sepsis, bacteremia, community-acquired pneumonia, UTIs, and intra-abdominal infections, with other localized infections included if sequential data are available. Immunocompromised patients should be analyzed separately for sensitivity analysis. The research should examine daily CBC measurements and vital signs recorded every 6-12 hours over the first 3-7 days of antibiotic therapy to capture early treatment response patterns. Clinical improvement should be defined by normalization of vital signs and CBC indices, shorter hospital length of stay, reduced ICU transfers, clinical resolution, and avoidance of antibiotic escalation. Please prioritize peer-reviewed medical literature, clinical studies, and original research publications from reputable medical journals, with preference for studies that include longitudinal data analysis and real-time monitoring approaches in hospital settings.\newline

\noindent\textbf{Clinical Data Provided by User:}
Clinical data is provided by user with selected training features (1-36, 39-42) and target feature as infection (43) from Table 4.\newline

\noindent\textbf{Brief-2 by Data Analysis Agent:}
The below are found to be major features and their importance scores to predict the target feature infection from data analysis tool:\newline
\noindent
t0\_temp\_avg\_squared: $0.4720$,\\
t0\_temp\_avg\_sqrt: $0.2752$,\\
mono: $0.0796$,\\
neut103uL: $0.0307$,\\
ig103uL: $0.0267$,\\
ig\_total\_neu\_ratio: $0.0263$,\\
t0\_hr\_avg\_squared: $0.0259$,\\
lymph: $0.0258$,\\
ret106uL: $0.0209$,\\
t0\_hr\_avg\_sqrt: $0.0168$\newline

\noindent\textbf{Reflection by Research Supervisor Agent:}
Based on BRIEF-1 and BRIEF-2, I need to generate 5 specific research topics that address the user's research question about sequential CBC and vital sign measurements as indicators of antibiotic treatment efficacy in bacterial infections.  Key elements from BRIEF-1:\newline
- Focus on sequential measurements of CBC features (neutrophil \%, eosinophil \%, immature granulocyte \%)\newline
- Vital sign trends (temperature, heart rate, respiratory rate, MAP, oxygen saturation)\newline
- Real-time indicators of antibiotic treatment efficacy and clinical improvement\newline
- Hospitalized adult patients with bacterial infections (sepsis, bacteremia, pneumonia, UTIs, intra-abdominal)\newline
- Daily CBC measurements and vital signs every 6-12 hours over first 3-7 days of antibiotic therapy.\newline
Key biomarkers from BRIEF-2 with importance scores:\newline
\noindent
t0\_temp\_avg\_squared: $0.4720$ (highest)\\
t0\_temp\_avg\_sqrt: $0.2752$\\
mono: $0.0796$\\
neut103uL: $0.0307$\\
ig103uL: $0.0267$\\
ig\_total\_neu\_ratio: $0.0263$\\
t0\_hr\_avg\_squared: $0.0259$\\
lymph: $0.0258$\\
ret106uL: $0.0209$\\
t0\_hr\_avg\_sqrt: $0.0168$\newline
I need to create 5 topics that cover:\newline
1. Temperature trends as predictors of treatment response (highest importance score)\newline
2. Heart rate variability and treatment efficacy monitoring\newline
3. Neutrophil and immature granulocyte dynamics during antibiotic therapy\newline
4. Monocyte and lymphocyte changes as treatment response indicators\newline 
5. Combined CBC and vital sign patterns for real-time treatment monitoring. Each topic should be specific, independently researchable, and contribute to the overall research question about sequential monitoring for antibiotic efficacy\newline

\noindent\textbf{Topic Assignments for Research Sub-Agents:}\newline
1)	Temperature trend patterns as predictors of antibiotic treatment response in hospitalized patients with bacterial infections.\newline
2)	Heart rate variability and trends as real-time indicators of antibiotic treatment efficacy in bacterial infections.\newline
3)	Neutrophil and immature granulocyte dynamics during antibiotic therapy as treatment response biomarkers.\newline
4)	Monocyte and lymphocyte count changes as indicators of antibiotic treatment response in bacterial infections.\newline
5)	Combined CBC and vital sign monitoring patterns for real-time assessment of antibiotic treatment efficacy.\newline

\noindent\textbf{Completion Summary by Reporting Agent:}\newline
Based on comprehensive research across 5 specific topics, I have gathered substantial evidence on sequential CBC and vital sign measurements as indicators of antibiotic treatment efficacy in bacterial infections. The research covered temperature trend patterns (highest predictive importance at 0.4720), heart rate variability monitoring, neutrophil and immature granulocyte dynamics, monocyte and lymphocyte changes, and combined multiparameter monitoring approaches.    Comprehensive Evidence for Sequential Monitoring of Antibiotic Treatment Efficacy\newline
1. Temperature Patterns (t0\_temp\_avg\_squared: $0.4720$ \\
 - highest predictive value) Clinical Evidence:\newline
-  Temperature trend analysis can predict hospital-acquired sepsis 24-48 hours before clinical diagnosis, providing a substantial time window for early intervention\newline
-  Median time advantage:  Temperature abnormalities occur 9 hours before first culture order and 17 hours before antibiotic administration\newline
-  Clinical Utility:  Sequential temperature monitoring every 6 hours may delay diagnosis by several hours compared to continuous monitoring, suggesting need for more frequent assessment\newline 
-  Thermoregulatory Response:  No statistical differences found between Gram-negative vs Gram-positive infections or septic shock vs non-shock patients in temperature patterns\newline
2. Heart Rate Variability (t0\_hr\_avg\_squared: $0.0259$\\
, t0\_hr\_avg\_sqrt: $0.0168$)   Clinical Evidence:\newline 
-  Early Detection Capability:  Continuous multi-parameter heart rate variability analysis can herald onset of sepsis before conventional clinical diagnosis\newline
-  Sensitivity Advantage:  HRV analysis is more sensitive at identifying sepsis than conventional vital signs, particularly in neonatal populations\newline
-  Monitoring Methodology:  HRV can be characterized continuously using time, frequency, complexity, and scale-invariant domain techniques\newline
-  Clinical Implementation:  Baseline HRV defined as mean variability for first 24 hours, with percentage changes tracked relative to clinical diagnosis timing\newline
3. Neutrophil and Immature Granulocyte Dynamics (neut103uL: $0.0307$, ig103uL: $0.0267$, ig\_total\_neu\_ratio: $0.0263$)   Clinical Evidence:\newline 
-  Immature Granulocyte Monitoring:  ROC analysis for treatment adequacy showed AUC of $0.692$ (95\% CI: 0.58-0.805) with sensitivity 73.5\% and specificity 57.9\% using IG\% cutoff of 0.75 on day 3\newline
- Treatment Response Assessment:  IG ratio demonstrated AUC of 0.676 with sensitivity 83.7\% and specificity 52.6\% using cutoff of 0.915\newline 
-   Clinical Utility:  IGs are automatically measured during routine CBC without additional sample requirements, available 24/7 with objective numerical results \newline
-  Neutrophil CD64 Index:  Superior performance demonstrated for diagnosing, monitoring bacterial infection, and evaluating antibiotic therapy with significant decline indicating effective treatment\newline
4. Monocyte and Lymphocyte Changes (mono: $0.0796$, lymph: $0.0258$)   Clinical Evidence:\newline
-  Monocyte-to-Lymphocyte Ratio (MLR):  Predictive role established in bacterial infections with NEWS score $\geq 7$ used for severe infection stratification\newline
-  Lymphocytopenia Significance:  Better predictor of bacteremia than conventional infection markers like temperature and WBC count\newline
-  Clinical Application:  Absolute monocyte count divided by absolute lymphocyte count provides objective assessment of immune status during treatment\newline
5. Combined CBC and Vital Sign Monitoring   Clinical Evidence:\newline
-  Cell Population Data (CPD) Integration:  Five key parameters (fluorescence intensity neutrophils, hemoglobin content of reticulocytes, absolute neutrophils count, antibody secreting lymphocytes, absolute immature granulocytes count) can be collected from same CBC sample\newline 
-  Cost-Effectiveness:  CPD parameters available at lower cost compared to CRP and PCT while providing comparable predictive value\newline
-  Intensive Care Infection Score (ICIS):  Increases in critically ill patients with probable infection, potentially useful predictor calculated from routine CBC data\newline
Quantitative Performance Metrics  | Biomarker | AUROC | Sensitivity | Specificity | Cutoff Value | Clinical Context | |-----------|-------|-------------|-------------|--------------|------------------| | IG\% (Day 3) | 0.692 | 73.5\% | 57.9\% | 0.75\% | Treatment adequacy | | IG Ratio | 0.676 | 83.7\% | 52.6\% | 0.915 | Treatment response | | Temperature Trends | N/A | N/A | N/A | N/A | 24-48h early detection | | HRV Analysis | N/A | Superior to conventional | N/A | N/A | Continuous monitoring |  Critical Analysis Linking to Data Features   Temperature Dominance:  The highest importance scores for temperature-related features (0.4720 and 0.2752) align with clinical evidence showing temperature patterns can predict sepsis 24-48 hours before conventional diagnosis. The mathematical transformations (squared and square root) likely capture non-linear temperature dynamics during infection progression and treatment response.   Moderate CBC Feature Importance:  The moderate importance scores for neutrophil (0.0307), immature granulocyte (0.0267), and lymphocyte (0.0258) parameters correspond with clinical evidence showing these markers provide objective, automated assessment but with moderate diagnostic accuracy when used individually.   Heart Rate Secondary Role:  Lower importance scores for heart rate features (0.0259, 0.0168) suggest these parameters provide complementary rather than primary diagnostic information, consistent with literature showing HRV as sensitive but requiring continuous monitoring for optimal utility.\newline
Remaining Gaps\newline
-  Limited Sequential Validation:  Most studies focus on single time-point measurements rather than sequential monitoring over 3-7 days of antibiotic therapy\newline
-  Treatment Response Kinetics:  Insufficient data on optimal timing intervals for CBC and vital sign measurements during antibiotic treatment 
-  Pathogen-Specific Patterns:  Limited evidence for differential monitoring approaches based on specific bacterial pathogens or infection sites\newline
-  Combined Score Validation:  No validated composite scores integrating temperature, heart rate, and CBC parameters for real-time treatment efficacy assessment.\newline
Clinical Implementation Recommendations   Optimal Monitoring Protocol:\newline
1.  Temperature:  Continuous or every 2-3 hours rather than standard 6-hour intervals\newline
2.  CBC Parameters:  Daily measurements focusing on IG\%, neutrophil counts, and lymphocyte trends\newline
3.  Heart Rate:  Continuous HRV monitoring when available, particularly in ICU settings \newline
4.  Combined Assessment:  Integration of temperature trends with CBC dynamics for treatment response evaluation\newline
Early Warning Thresholds:\newline
- Temperature pattern changes 24-48 hours before clinical deterioration - IG\% >0.75 on day 3 suggesting inadequate treatment response\newline
- Declining lymphocyte counts indicating persistent immune suppression\newline
- Loss of HRV complexity suggesting ongoing sepsis progression.
The evidence strongly supports the clinical utility of sequential CBC and vital sign monitoring for antibiotic treatment efficacy assessment, with temperature patterns showing the highest predictive value and CBC parameters providing objective, cost-effective complementary information for real-time clinical decision support.\newline

\noindent\textbf{Variable Importance and Explanation Plots:}\par
\vspace{0.3em}

% --- Figure 3 ---
\begin{figure}[!htbp]
  \centering
  \includegraphics[width=0.75\linewidth]{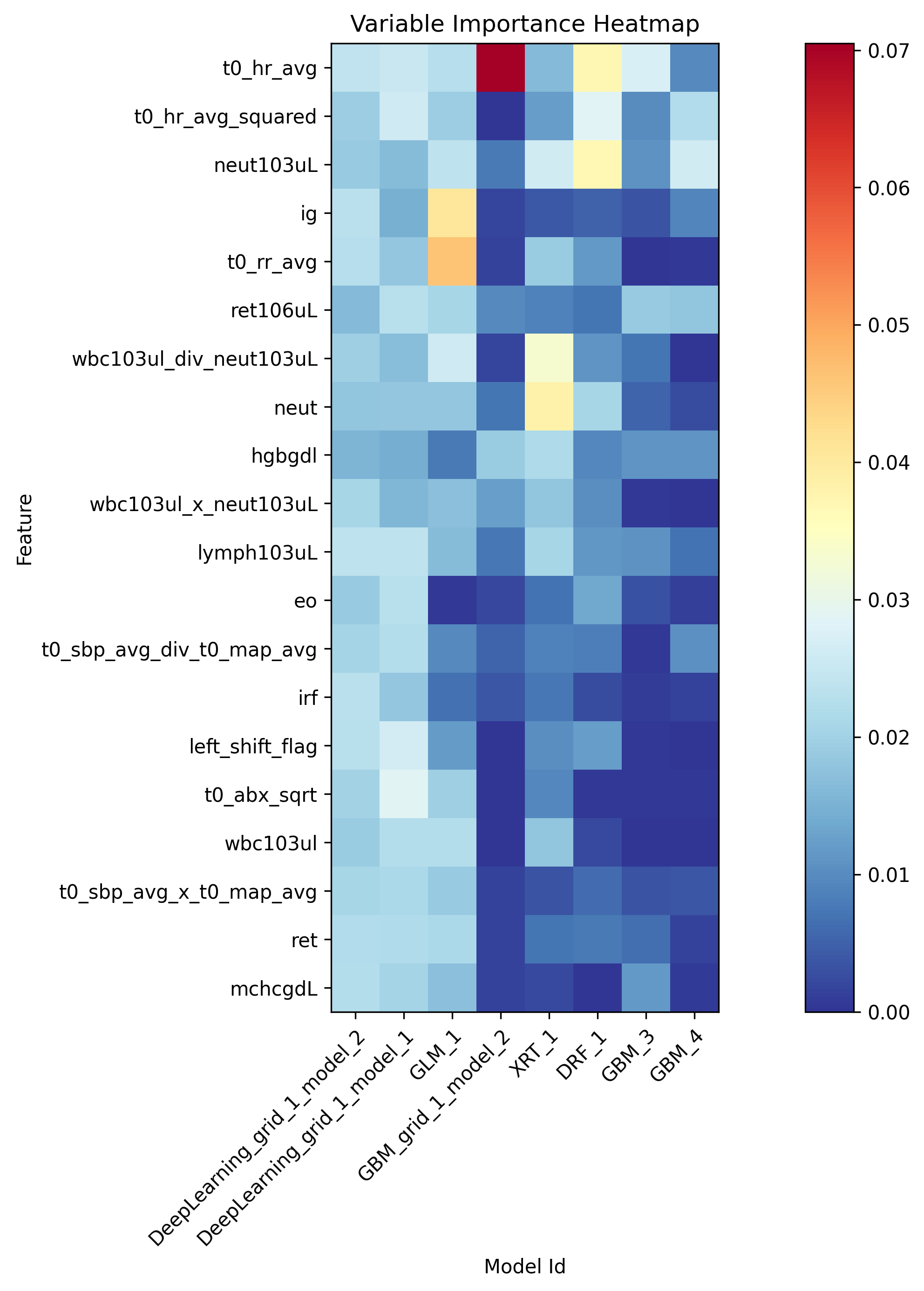}
  \caption{Heatmap highlighting the feature importances across multiple trained models by the Data Analysis Agent.}
  \label{fig:heatmap}
\end{figure}
\FloatBarrier  % stops fig.3 from jumping

% --- Figure 4 ---
\begin{figure}[!htbp]
  \centering
  \includegraphics[width=0.75\linewidth]{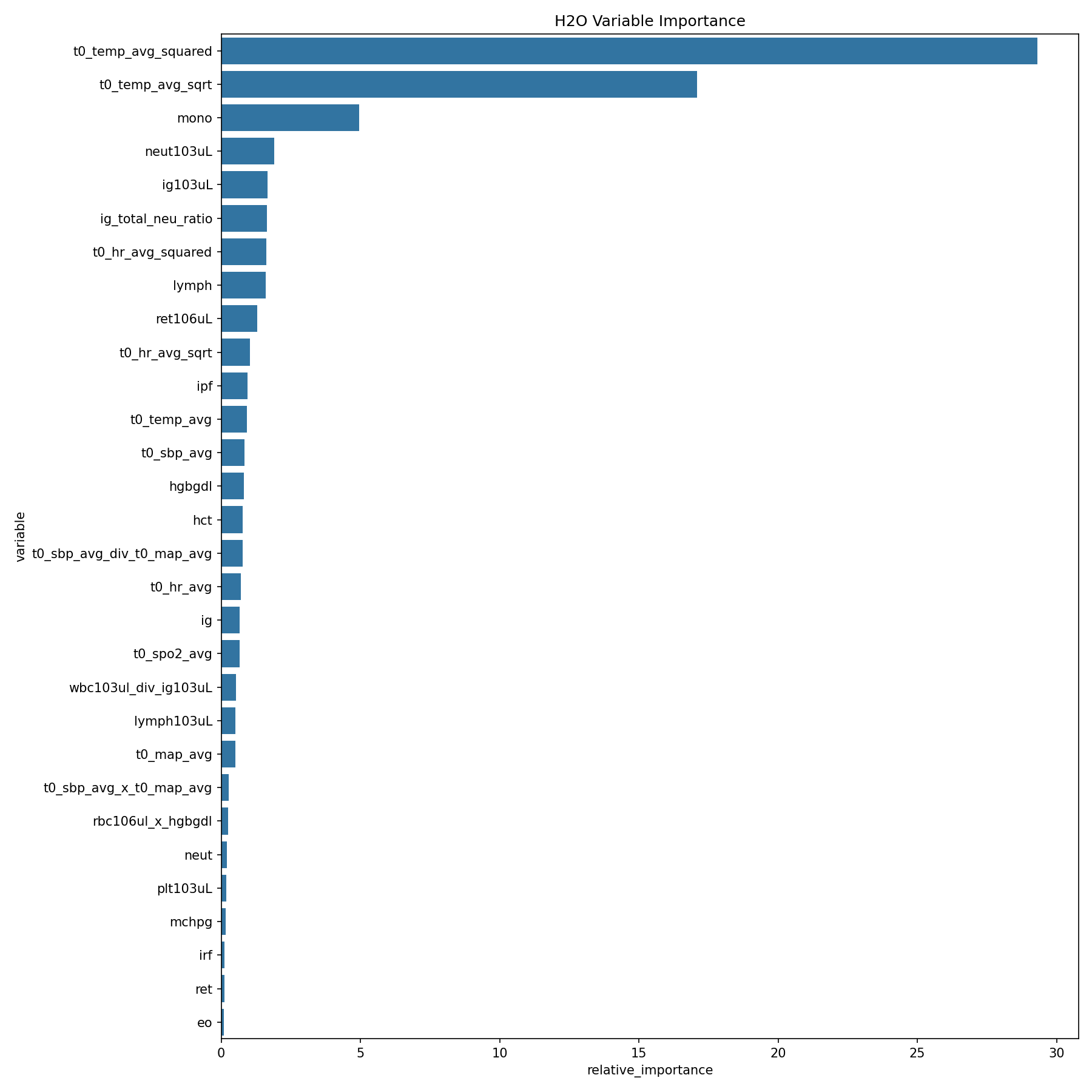}
  \caption{Identified features and their respective relative importances obtained from the best performing model, Gradient Boost Machine (GBM), achieving a root mean square error of 0.2065.}
  \label{fig:barplot}
\end{figure}
\FloatBarrier  % stops fig.4 from overtaking fig.5

% --- Figure 5 ---
\begin{figure}[H]
  \centering
  \includegraphics[width=0.75\linewidth]{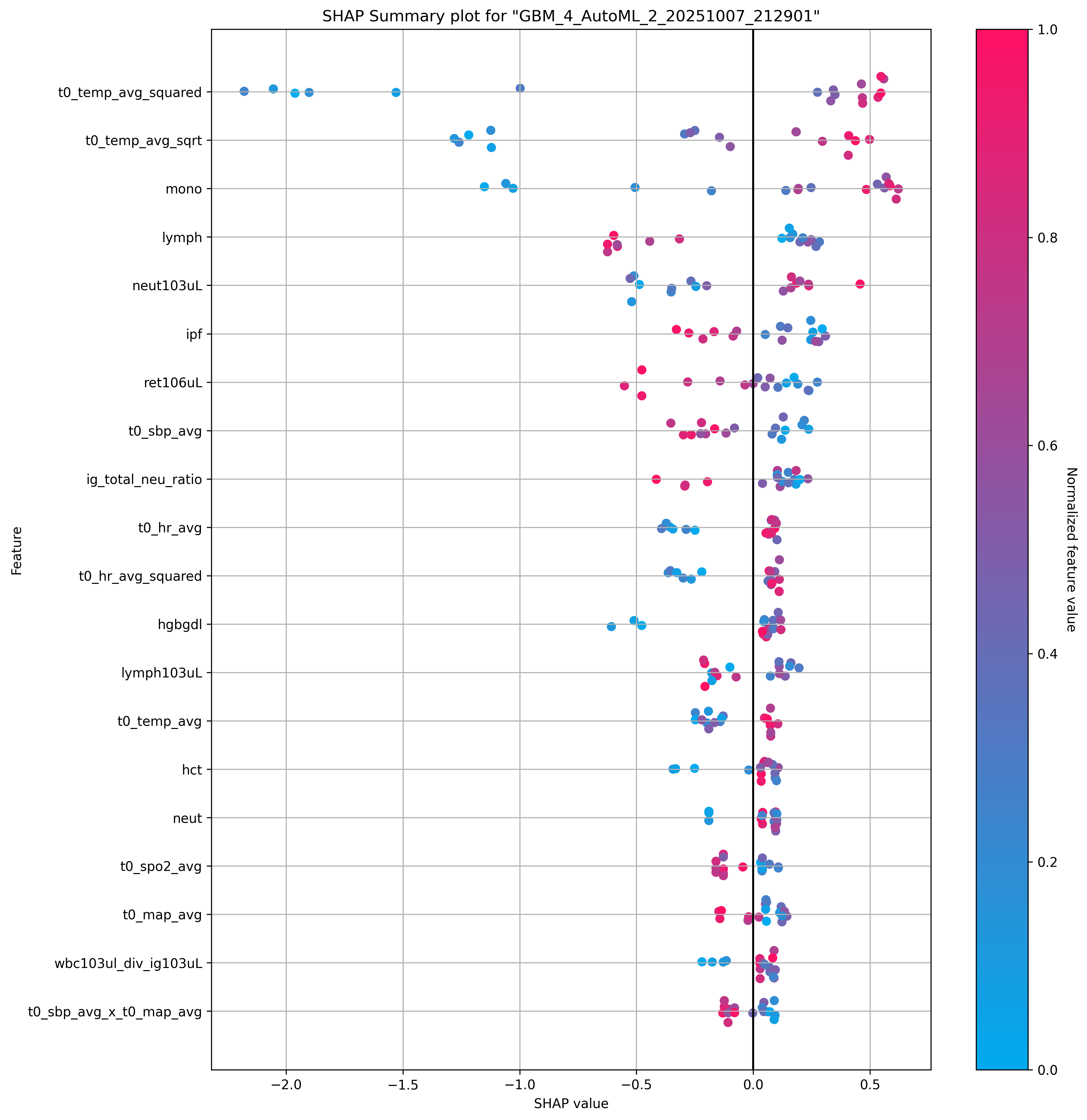}
  \caption{The SHAP explanations showing feature importance and normalized feature-value distributions for infection prediction by the GBM model.}
  \label{fig:shap}
\end{figure}
\FloatBarrier

\end{document}